\documentclass[a4paper]{spie}  

\usepackage[dvips]{graphicx}

\title{Measurement of Coupling PDC photon sources with single-mode and multimode optical fibers}

\author{ Stefania Castelletto\supit{b}, Ivo Pietro Degiovanni\supit{b}, Alan Migdall\supit{a}, Valentina Schettini\supit{b}, Michael
Ware\supit{a}\\
 \supit{a}Optical Technology Division,\\
 National Institute of Standards
and Technology,Gaithersburg, Maryland 20899-8441,\\
 \supit{b}Istituto Elettrotecnico Nazionale G.
Ferraris,
Strada delle Cacce 91-10135 Torino (Italy) \\
}

\authorinfo{Further author information: (Send
correspondence to Stefania Castelletto)\\
E-mail: castelle@ien.it, Telephone: 0039 011 3919223}

\begin{document}

 \maketitle

\begin{abstract}
We investigate the coupling efficiency of parametric
downconversion light (PDC) into single and multi-mode optical
fibers as a function of the pump beam diameter, crystal length and
walk-off. We outline two different theoretical models for the
preparation and collection of either single-mode or multi-mode PDC
light (defined by, for instance, multi-mode fibers or apertures,
corresponding to bucket detection). Moreover, we define the
mode-matching collection efficiency, important for realizing a
single-photon source based on PDC output into a well-defined
single spatial mode. We also define a multimode collection
efficiency that is useful for single-photon detector calibration
applications.

\end{abstract}

\section{Introduction} Earliest studies of parametric
downconversion (PDC) addressed problems in fundamental physics,
while  more recent studies target applications such as quantum
metrology \cite{MCD02} and quantum information \cite{TBZ00,KLM01}.
While both of these areas make use of two-photon light, they are
distinct applications that present different requirements for that
light. The stringent requirements of these applications are
driving researchers to optimize the PDC process. For these efforts
to succeed, a clear theoretical framework is needed.

PDC produces a quantum state of light with a two-photon field
description. However, if only one photon of the pair is measured,
the source exhibits purely thermal statistical behavior, given by
intrinsic multi-mode photon production. We can however introduce a
certain degree of coherence. By measuring one of the photons, we
prepare the other photon in a specific state. The prepared state
will be pure only if we project the first PDC photon (called also
the heralding photon) into a pure state. In each of the above PDC
applications, we prepare one photon by measuring its twin. Thus
for optimization of the process, it is crucial to have a proper
definition and measurement of the efficiency of that preparation
and of the related mode-matching. We present two different models
to define and optimize the two-photon-mode preparation and
mode-matching efficiencies. They are distinguished by how the
heralding photon is collected. One uses a multimode spatial filter
or a bucket detection system, while the second uses a single-mode
fiber. A bucket detector is a multimode detector where all the
modes are detected jointly. Hence the information about the
location of the detected photon (or equivalently in what mode the
photons were detected) is "erased". We obtain different
dependencies of the efficiency on the pump parameters in these two
arrangements. This is particularly important for two specific
applications: the calibration of a single-photon detector and the
realization of a single-photon on demand source (SPOD)
\cite{PJF02,MBC02}.

\section{Biphoton field}
 We consider a two-photon wavefunction,
written as \cite{RUB96}
\begin{equation} \label{psirt}
      \left| \psi \right> =
       \
      \int\textrm{d}^{2}\rho_{1} \textrm{d}^{2}\rho_{2} \textrm{d}
      t_{1}  \textrm{d}
      t_{2}
\widetilde{\Phi}(\mathbf{\rho}_{1},\mathbf{\rho}_{2},t_{1},t_{2})|1_{\mathbf{\rho}_\mathrm{2},t_\mathrm{2}}
      \rangle |1_{\mathbf{\rho}_\mathrm{1},t_\mathrm{1}}
      \rangle,
\end{equation}
where $\mathbf{\rho}_{1,2}$ represents the transverse positions of
the two photons at the instant $t_{1,2}$ and
$\widetilde{\Phi}(\mathbf{\rho}_{1},\mathbf{\rho}_{2},t_{1},t_{2})$
is the biphoton field. Its calculation is analytic only for
first-order approximation of the transverse wavevectors (assuming
the pump, signal, and idler have narrow transverse angular
distributions we can adopt the paraxial approximation), and is
obtained by performing the Fourier transform of the pump angular
distribution and the phase-matching function with respect to the
pump transverse k-components and the signal k-components. Perfect
transverse phase-matching is also assumed. The result derived in
ref. \cite{CDM04} is
\begin{eqnarray}
       \nonumber
\widetilde{\Phi}(\mathbf{\rho}_{1},\mathbf{\rho}_{2},t_{1},t_{2})&=&N_{1}
       \exp[\frac{-i (K_{i} \theta_{i}^{2} + K_{s}
       \theta_{s} \theta_{i} \tau)}{D}]
        \exp[\frac{-(\mathcal{N}_{p}-\mathcal{N}_{s})^{2}
\tau^{2}}{D^{2} w_{p}^{2}
        K_{p}}]\\ \nonumber & &
        \times \exp[\frac{2(\mathcal{N}_{p}-\mathcal{N}_{s}) \tau
(y_{1}+\frac{\theta_{i} \tau}{D})}{D w_{p}^{2}
        K_{p}}]
        \exp[-\frac{x_{1}^{2}+ (y_{1}+\frac{\theta_{i} \tau}{D})^{2}}{
        w_{p}^{2}}] \Pi_{D
        L}(\tau)\nonumber \\ & &
         \times \delta(x_{1}-x_{2})
\delta(y_{1}-y_{2}+\frac{(\theta_{i}+\theta_{s}) \tau}{D})
,\label{phirt}
\end{eqnarray}
where $\tau=t_{1}-t_{2}$ and $\Pi_{D L}(\tau)=1$ for
$0\leq\tau\leq D L$ and $0$ elsewhere.  The subscripts $s$, $i$,
$p$ indicate the signal, idler, and pump. $\theta_{i,s}$ are the
central emission angles (in a small angle non-collinear
approximation), and
$K_{i,s,p}=n_{i,s,p}(\Omega_{i,s,p},\phi)\Omega_{i,s,p}/c$
describe the directions of the central intensities of the
wavevectors. The terms,
$\mathcal{N}_{p}=\frac{\Omega_{p}}{c}\frac{\mathrm{d}n_{p}(\Omega_{p},\phi)}{\mathrm{d}\phi}|_{\phi_{o}}$
and
$\mathcal{N}_{s}=\frac{\Omega_{s}}{c}\frac{\mathrm{d}n_{s}(\Omega_{s},\phi)}{\mathrm{d}\phi}|_{\phi_{o}}$
account for the effects on the refractive indexes (
$n_{p,s,i}(\omega_{p,s,i}, \phi )$ expanded around the central
frequencies ($\Omega_{s,i}$), and around the phase-matching angle
$\phi_{o}$) of the pump and the signal due to the pump angular
spread, which are responsible for a small deviation from the
phase-matching angle $\phi_{o}$.  $D =
       -\frac{\textrm{d}
n_\mathrm{i}(\omega_\mathrm{i})\omega_\mathrm{i}/c}{\textrm{d}\omega_\mathrm{i}}
      |_{\Omega_\mathrm{i}}
      + \frac{\textrm{d} n_\mathrm{s}(\omega_\mathrm{s},\phi
)\omega_\mathrm{s}/c}{\textrm{d}\omega_\mathrm{s}}
          |_{\Omega_\mathrm{s}}$  is
the differential phase velocity between the signal and idler
photons in the crystal. $D=0$ and $\mathcal{N}_{s}=0$ for type I
degenerate phase-matching. The pump-beam transverse field
distribution is assumed to be Gaussian with a waist of
$w_\mathrm{p}$ at the crystal. We also assume that the pump
propagates with negligible diffraction inside the crystal. In the
following, the idler and signal space-time positions are indicated
by 1 and 2, respectively.
\section{Single mode efficiency}
\begin{figure}[tbp]
\par
\begin{center}
\includegraphics[angle=0, width=12 cm, height=9 cm]{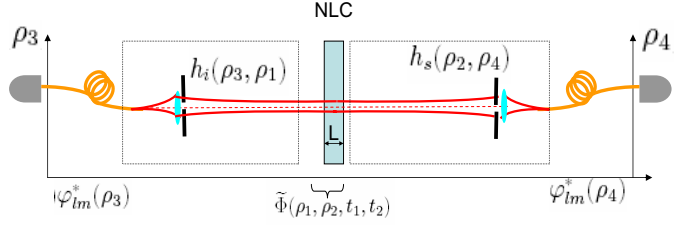}
\end{center}
\caption{Unfolded picture of an object  corresponding to the idler
channel of a PDC source,  generated in a nonlinear crystal NLC of
length $L$. The object (on the left) is a coherent source when the
idler beam is prepared by single-mode fiber. \label{fig1}}
\end{figure}

Envisioning the optics setup as the unfolded scheme of
Fig.~\ref{fig1} \cite{DNK88}, the source is described by the
propagation of a coherent mode defined by
$\varphi_{lm}(\mathbf{\rho}_3 ) $ through an optical element with
impulse response function $h_{i}(\mathbf{\rho}_3 ,\mathbf{\rho}_1
)$, through the non-linear crystal where the mode gets transformed
according to the phase matching function $
\widetilde{\Phi}(\mathbf{\rho}_{1},\mathbf{\rho}_{2},t_{1},t_{2})$,
and collected eventually by $h_{s}(\mathbf{\rho}_2
,\mathbf{\rho}_4  )$. The actual collected mode will then be given
by the field $\varphi_{lm}(\mathbf{\rho}_4 )$.

The coincidences measured at the positions 3 and 4 are then
 $\mathcal{C}_{34}$, the overlap between the PDC field and both of
 the preparation or collection modes, while the single counts
$\mathcal{C}_3$ and $\mathcal{C}_4$ measure  the overlap
individually between the biphoton field and each of the preparing
or collecting modes:
\begin{eqnarray}\label{eq:3}
      \mathcal{C}_{34}  &=& \int \textrm{d} t_1
       \textrm{d} t_2  \\ & & \nonumber \left|   \int
    \textrm{d}^2
    \rho_1 \textrm{d}^2
    \rho_2 \textrm{d}^2
    \rho_3 \textrm{d}^2
    \rho_4
    \widetilde{\Phi}( \mathbf{\rho}_1,t_{1}, \mathbf{\rho}_2,t_{2})
h_{i}(\mathbf{\rho}_1,\mathbf{\rho}_3)h_{s}(\mathbf{\rho}_2,\mathbf{\rho}_4)
\varphi^{*}_{l'm'} (\rho_{4}) \varphi^{*}_{lm} ( \rho_{3})
\right|^2
\\ \nonumber
      \mathcal{C}_{3}  &=& \int  \textrm{d} t_1 \textrm{d}
t_2 \textrm{d}^2
    \rho_4 \left|  \textrm{d}^2
    \rho_1 \textrm{d}^2
    \rho_2 \textrm{d}^2
    \rho_3
       ~  \widetilde{\Phi}( \mathbf{\rho}_1,t_{1}, \mathbf{\rho}_2,t_{2})
h_{i}(\mathbf{\rho}_1,\mathbf{\rho}_3)h_{s}(\mathbf{\rho}_2,\mathbf{\rho}_4)
\varphi^{*}_{lm} ( \rho_{3})
   \right|^2 \\ \nonumber
         \mathcal{C}_{4}  &=& \int  \textrm{d} t_1 \textrm{d}
t_2 \textrm{d}^2
    \rho_3 \left|  \textrm{d}^2
    \rho_1 \textrm{d}^2
    \rho_2 \textrm{d}^2
    \rho_3
       ~  \widetilde{\Phi}( \mathbf{\rho}_1,t_{1}, \mathbf{\rho}_2,t_{2})
h_{i}(\mathbf{\rho}_1,\mathbf{\rho}_3)
h_{s}(\mathbf{\rho}_2,\mathbf{\rho}_4)\varphi^{*}_{lm} ( \rho_{4})
   \right|^2
   \end{eqnarray}

\begin{figure}[tbp]
\par
\begin{center}
\includegraphics[angle=0, width=8cm, keepaspectratio=true]{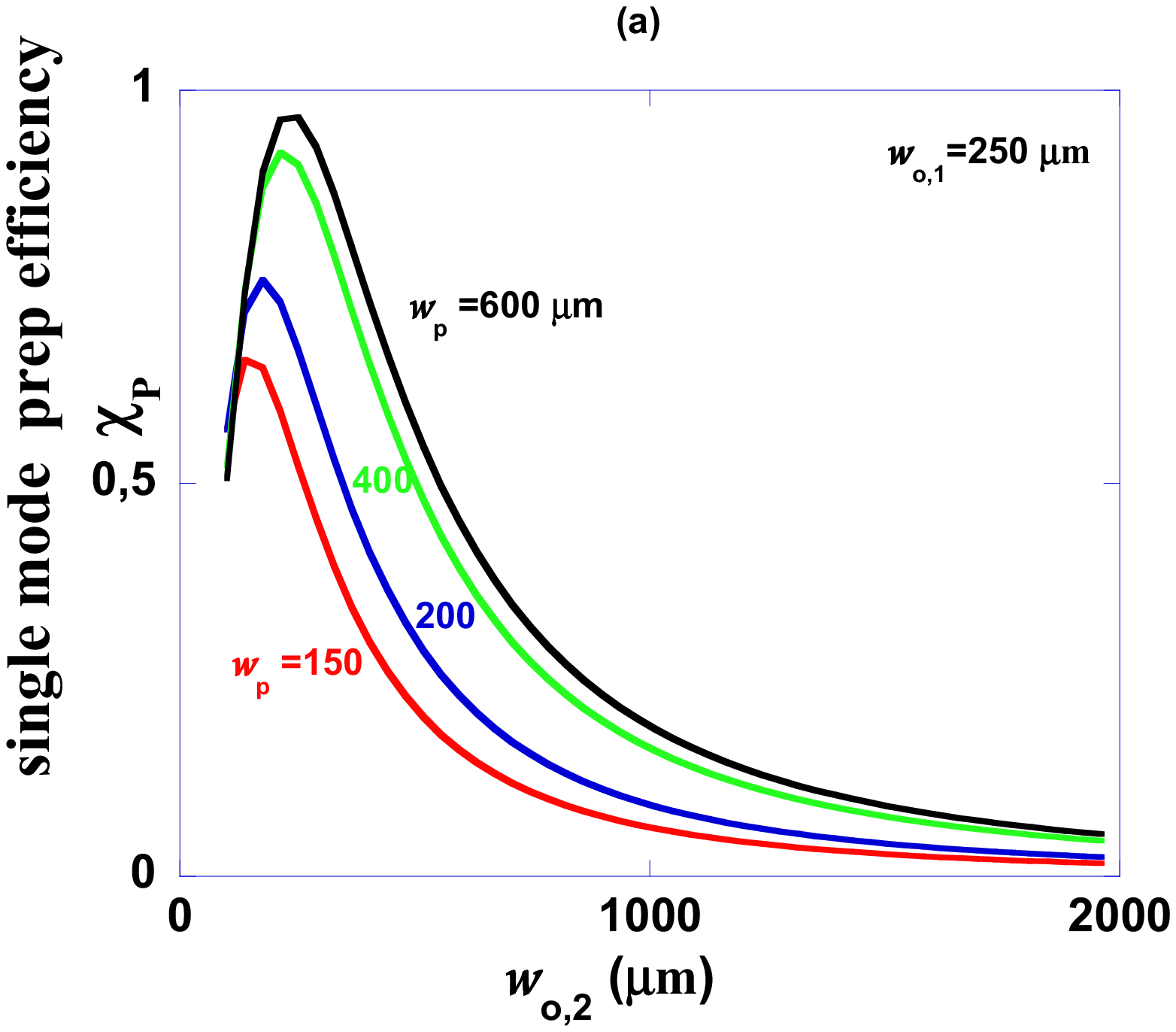}
\includegraphics[angle=0, width=8cm, keepaspectratio=true]{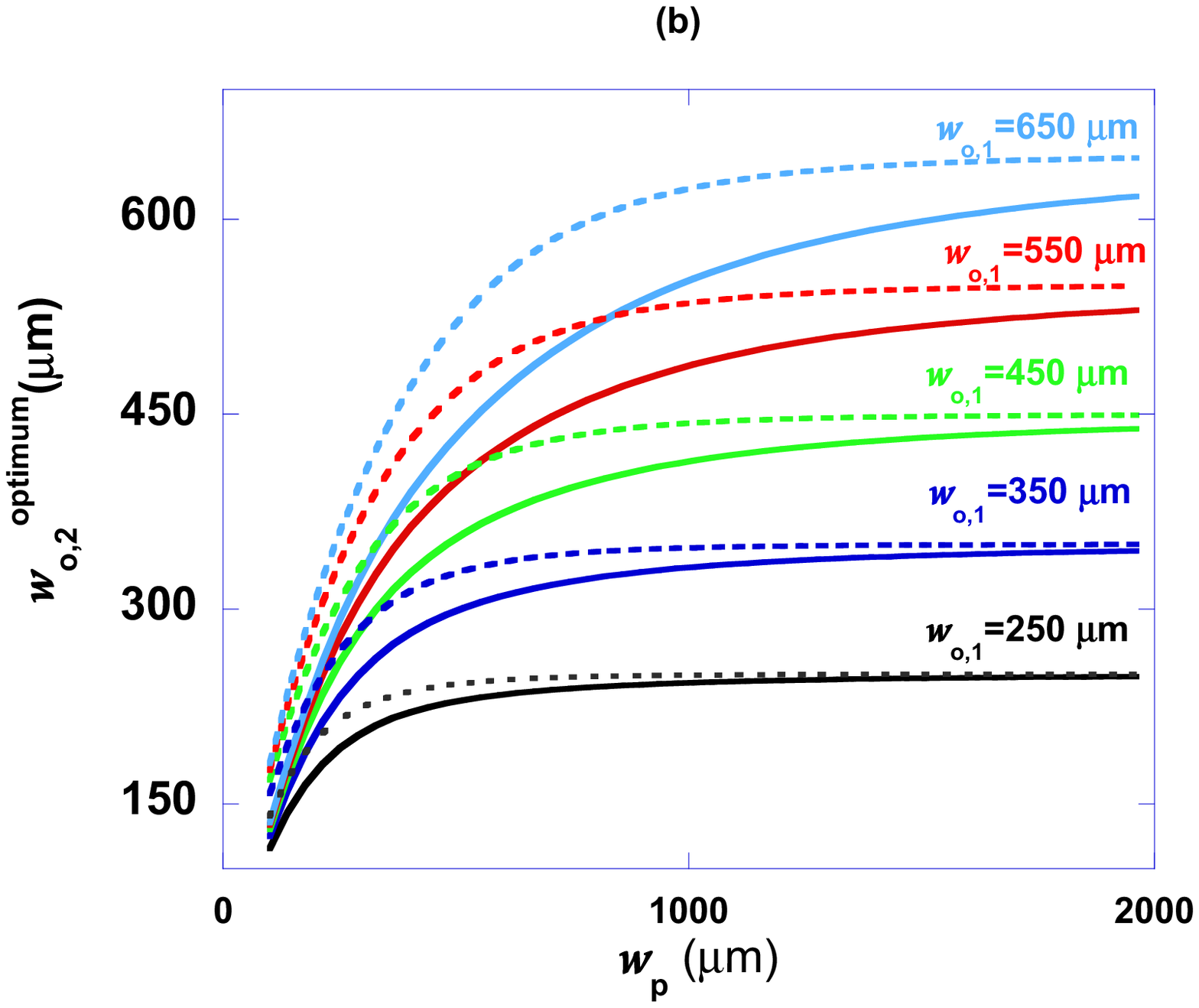}
\end{center}
\caption{(a) Plot of $\chi_{P}$ versus the collecting mode waist
for various pump waist  $w_{p}=150, 200,400,600$ $\mu$m at
$w_{o,1}=250$ $\mu$m (b) Plot of the optimum waist $w_{o,2}$
versus the pump waist for various preparing waists $w_{o,1}$ in
the case of $\chi_{M}$ (dashed line) and $\chi_{P}$ (solid line).
\label{fig2}}
\end{figure}

Here we distinguish between the single-mode matching efficiency
calculated by
\begin{equation}\label{chiM}
   \chi_{M} = \frac{\mathcal{C}_{34}}{ \sqrt{\mathcal{C}_3 \mathcal{C}_4}
   },
\end{equation}
and the single-mode preparation efficiency
\begin{equation}\label{chiP}
   {\chi_{P}}^{(3,4)} = \frac{\mathcal{C}_{34}}{ \mathcal{C}_{3,4}  }.
\end{equation}
Eq.(\ref{chiM}) is the appropriate efficiency to optimize when the
final goal is a single photon source on demand emitted in defined
photon modes, while Eq.(\ref{chiP}) is the efficiency measured
when single photon detector calibration is required. Specifically
in the paper, we calculate the efficiency in a perfect imaging
configuration, namely
$h_s(\mathbf{\rho}_2,\mathbf{\rho}_4)=\delta( \mathbf{ \rho}_2-M_4
\mathbf{\rho}_4)$ and
$h_i(\mathbf{\rho}_3,\mathbf{\rho}_1)=\delta( \mathbf{\rho}_1-M_3
\mathbf{\rho}_3)$ (lenses have infinite aperture with
magnification $M_{j+2}$). The lenses are arranged to place the
preparation and collection beam waists, $w_{o,j}=250$ $\mu$m at
the crystal, with guided Gaussian field modes $ \varphi_{10}
(\mathbf{\rho}_{j+2}) = \sqrt{\frac{2}{\pi}}
\frac{M_{j+2}}{w_{o,j}}
      \exp \left[ - \frac{\rho_{j+2}^2 M_{j+2}^{2}}{ w_{o,j}^2}
      \right],$
with $j=1,2$. The spatial coherence of the single guided modes in
the signal and idler arms should ultimately match the overall
spatial coherence of the two-photon states. By explicit
calculation the single-mode matching efficiency is then given by
\begin{equation}\label{chi12}
\chi_{M}=\mathcal{F}_M \frac{4~ w_{o,1}^2 w_{o,2}^2
w_{p}^{2}~\sqrt{A}}{\sqrt{(w_{o,2}^{2}w_{p}^{2}+
w_{o,1}^{2}(w_{o,2}^{2}+ w_{p}^{2}))^{3} B}},
\end{equation}

with $A=(-\mathcal{N}_{p}+\mathcal{N}_{s}+K_{p} \theta_{i})
~(\mathcal{N}_{p}-\mathcal{N}_{s}+ K_{p}
\theta_{s})\sqrt{w_{o,1}^2+w_p^2} \sqrt{w_{o,2}^2+w_p^2}$,
\begin{equation}\label{F} \mathcal{F}_M=
 \frac{ \mathrm{Erf}[ \frac{L \sqrt{2 B}}{K_{p} \sqrt{w_{o,2}^{2}
 w_{p}^{2}+w_{o,1}^{2}(
w_{p}^{2}+w_{o,2}^{2})}}]}{\sqrt{\mathrm{Erf}[\sqrt{2}
L\frac{(-\mathcal{N}_{p}+ \mathcal{N}_{s}+ K_{p}
\theta_{i})}{K_{p} \sqrt{w_{o,1}^{2}+
w_{p}^{2}}}]~\mathrm{Erf}[\sqrt{2}
L\frac{(\mathcal{N}_{p}-\mathcal{N}_{s}+ K_{p} \theta_{s}) }{K_{p}
\sqrt{w_{o,2}^{2}+ w_{p}^{2}}}]} }
\end{equation}
and $B=( \mathcal{N}_{p}-  \mathcal{N}_{s} +K_{p}\theta_{s})^2
w_{o,1}^2+ (- \mathcal{N}_{p}+ \mathcal{N}_{s}+ K_{p}\theta_{i})^2
w_{o,2}^2+ K_{p}^{2} (\theta_{i}+\theta_{s})^2 w_{p}^{2}.$ In the
thin crystal limit this reduces to
\begin{equation}
\chi_{M}=\frac{4~ w_{p}^{2} w_{o,1}^{2} w_{o,2}^{2}
\sqrt{(w_{o,1}^{2}+w_{p}^{2})
(w_{o,2}^{2}+w_{p}^{2})}}{(w_{o,2}^{2} w_{p}^{2}+
w_{o,1}^{2}~(w_{o,2}^{2}+ w_{p}^{2}))^{2}}
\end{equation}
as first calculated in Refs.~\cite{BGS03,CDW03}. The single mode
preparation efficiency is given by
\begin{equation}\label{chi12bis}
\chi_{P}^{(3,4)}=\mathcal{F}_P \frac{4~(- \mathcal{N}_{p}+
\mathcal{N}_{s}+ K_{p}\theta_{i}) w_{o,1}^2 w_{o,2}^2
w_{p}^{2}~\sqrt{w_{o,1}^2+w_{p}^2}}{\sqrt{(w_{o,2}^{2}w_{p}^{2}+
w_{o,1}^{2}(w_{o,2}^{2}+ w_{p}^{2}))^{3} B}},
\end{equation}
with
\begin{equation}\label{F} \mathcal{F}_P=
 \frac{ \mathrm{Erf}[ \frac{L \sqrt{2 B}}{K_{p} \sqrt{w_{o,2}^{2}
 w_{p}^{2}+w_{o,1}^{2}(
w_{p}^{2}+w_{o,2}^{2})}}]}{\mathrm{Erf}[\sqrt{2}
L\frac{(-\mathcal{N}_{p}+ \mathcal{N}_{s}+ K_{p}
\theta_{i})}{K_{p} \sqrt{w_{o,1}^{2}+ w_{p}^{2}}}]}
\end{equation}
Also in this case we can calculate easily the thin crystal limit

\begin{equation}\label{chiP0}
\chi_{P}=\frac{4~ w_{p}^{2} w_{o,1}^{2} w_{o,2}^{2}
(w_{o,1}^{2}+w_{p}^{2}) }{(w_{o,2}^{2} w_{p}^{2}+
w_{o,1}^{2}~(w_{o,2}^{2}+ w_{p}^{2}))^{2}}
\end{equation}
Note that
\begin{equation}\label{chiM0}
\chi_M=\sqrt{\chi_P^{(3)} \chi_P^{(4)}}.
\end{equation}
In this case the information of the location of the prepared and
detected modes is included in the measurement. In
Fig.\ref{fig2}(a), we plot the theoretical prediction of $\chi_P$
for a 5 mm crystal of LiIO$_{3}$ versus the collecting waist
$w_{o,2}$ for fixed preparing waist $w_{o,1}=250$ $\mu$m at the
crystal, for different pump waist configurations. As it appears
there is an optimum collecting waist and the best efficiency is
obtained for the larger pump waist. $\chi_M$ behaves as $\chi_P$,
except that the highest efficiency is obtained when the preparing
and collecting waist are the same ($w_{o,1}$=$w_{o,2}$). This can
be clearly observed in Fig.\ref{fig2}(b)where the optimum
collecting waist is plotted versus the pump waist for different
prepared waist $w_{o,1}$. The optimum $w_{o,2}$ (giving the
highest efficiency) approaches $w_{o,1}$ as the pump waist
increases, and this approach is faster in the $\chi_{M}$ case, as
expected.

\section{Multi-mode or bucket detection efficiency}
\begin{figure}[tbp]
\par
\begin{center}
\includegraphics[angle=0, width=12 cm, height=9 cm]{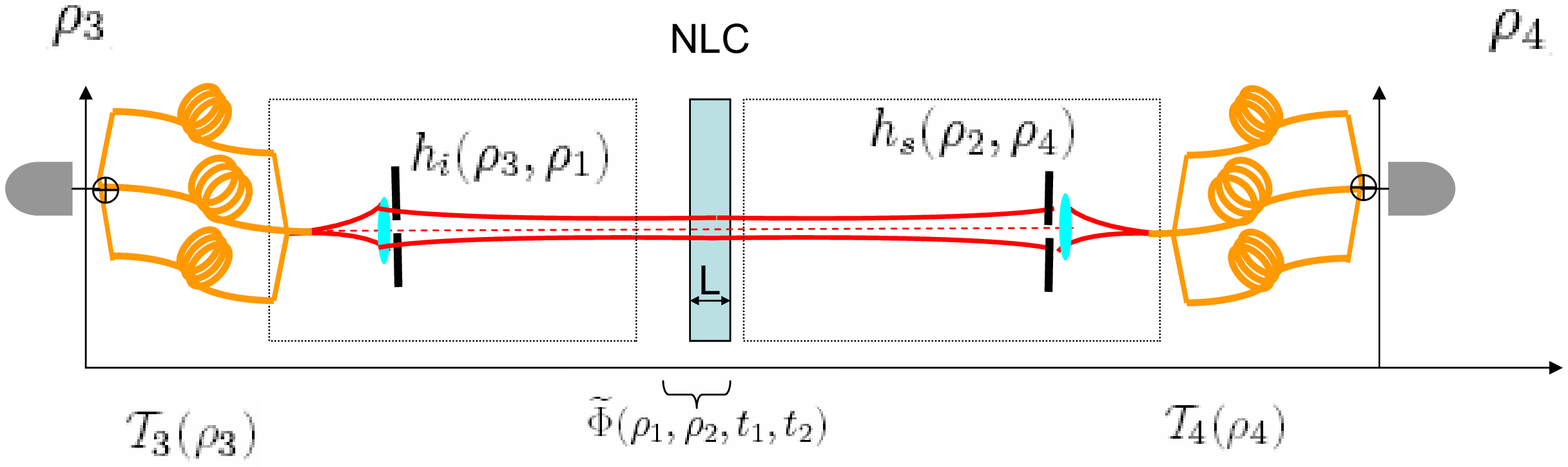}
\end{center}\caption{Unfolded picture of an object  corresponding to the idler
channel of a PDC source, generated in a nonlinear crystal NLC of
length $L$. The object (on the left) is an incoherent source
prepared by a multi-mode fiber, and the image is collected by a
multimode fiber. \label{fig3}}
\end{figure}

In the case of multi-mode or bucket detection efficiency the
schematic is Fig.~\ref{fig3}, where we overlap the PDC wave
function with the impulse response function of the optical system,
and eventually we average over the spatial distribution of
multi-mode fibers or spatial filters (apertures) or bucket
detector, erasing the information about the spatial location where
the photons where collected. More specifically, in Fig.
\ref{fig3}, the source is considered incoherent with finite
transverse distribution $\mathcal{T}_{3}(\mathbf{\rho}_3 ) $.
The collecting modes are then represented by the spatial filter
and described by $\mathcal{T}_4(\mathbf{\rho}_4 )$. The
coincidences measured at the positions 3 and 4 are then
 $C_{34}$, and $C_3$ and $C_4$ are the single counts:
\begin{eqnarray}
      C_{34}  &=& \int \textrm{d} t_1
       \textrm{d} t_2  \textrm{d}^2
    \rho_3 \textrm{d}^2
    \rho_4 \mathcal{T}_4(\rho_{4}) \mathcal{T}_3(\rho_{3})
    \\ & & \nonumber \left|   \int
    \textrm{d}^2
    \rho_1 \textrm{d}^2
    \rho_2 \widetilde{\Phi}( \mathbf{\rho}_1,t_{1}, \mathbf{\rho}_2,t_{2})
h_{i}(\mathbf{\rho}_1,\mathbf{\rho}_3)h_{s}(\mathbf{\rho}_2,\mathbf{\rho}_4)
 \right|^2
\\ \nonumber
      C_{3}  &=& \int  \textrm{d} t_1 \textrm{d}
t_2 \textrm{d}^2
    \rho_4 \textrm{d}^2
    \rho_3 \mathcal{T}_3( \rho_{3})\left|  \textrm{d}^2
    \rho_1 \textrm{d}^2
    \rho_2
       ~  \widetilde{\Phi}( \mathbf{\rho}_1,t_{1}, \mathbf{\rho}_2,t_{2})
h_{i}(\mathbf{\rho}_1,\mathbf{\rho}_3)h_{s}(\mathbf{\rho}_2,\mathbf{\rho}_4)
   \right|^2 \\ \nonumber
         C_{4}  &=& \int  \textrm{d} t_1 \textrm{d}
t_2 \textrm{d}^2
    \rho_3 \textrm{d}^2
    \rho_4 \mathcal{T}_4(\rho_{4})  \left|  \textrm{d}^2
    \rho_1 \textrm{d}^2
    \rho_2
       ~  \widetilde{\Phi}( \mathbf{\rho}_1,t_{1}, \mathbf{\rho}_2,t_{2})
h_{i}(\mathbf{\rho}_1,\mathbf{\rho}_3)
h_{s}(\mathbf{\rho}_2,\mathbf{\rho}_4)
   \right|^2
   \end{eqnarray}

The 2-photon multi-mode matching efficiency is then defined by
\cite{MRP98a}
\begin{equation}\label{eta}
\eta_{M}=\frac{C_{34}}{\sqrt{C_{3} C_{4}}}.
\end{equation}
 and the 2-photon multi-mode preparation efficiency is defined by
\begin{equation}
\eta_{P}^{(3,4)}=\frac{C_{34}}{C_{3,4} }.
\end{equation}

We calculate the multi-mode matching and preparation efficiencies,
assuming the incoherent source is completely determined by the
functions $
\mathcal{T}_{j+2}(\mathbf{\rho}_{j+2})=e^{-\frac{2\rho_{j+2}^2
M_{j+2}^2 }{w_{o,j}^2}} $, where the impulse response function is
the Dirac delta as before. Calculating the multi-mode matching
efficiency we obtain:
\begin{equation}\label{eta12}
\eta_{M}=\mathcal{F}_M \frac{ w_{o,1} w_{o,2}
~\sqrt{A}}{\sqrt{(w_{o,2}^{2}w_{p}^{2}+ w_{o,1}^{2}(w_{o,2}^{2}+
w_{p}^{2})) B}},
\end{equation}
In the thin crystal limit $\eta_{M}$ becomes
\begin{equation}
\eta_{M}=\frac{  w_{o,1}^{2} w_{o,2}^{2}
\sqrt{(w_{o,1}^{2}+w_{p}^{2})
(w_{o,2}^{2}+w_{p}^{2})}}{w_{o,2}^{2} w_{p}^{2}+
w_{o,1}^{2}~(w_{o,2}^{2}+ w_{p}^{2})}
\end{equation}
Also in this case the multi mode preparation efficiency is given
by
\begin{equation}\label{eta12bis}
\eta_{P}^{(3,4)}=\mathcal{F}_P \frac{4~(- \mathcal{N}_{p}+
\mathcal{N}_{s}+ K_{p}\theta_{i})  w_{o,2}^2
~\sqrt{w_{o,1}^2+w_{p}^2}}{\sqrt{(w_{o,2}^{2}w_{p}^{2}+
w_{o,1}^{2}(w_{o,2}^{2}+ w_{p}^{2})) B}},
\end{equation}
with thin crystal limit
\begin{equation}
\eta_{P}=\frac{  w_{o,1}^{2} w_{o,2}^{2}
(w_{o,1}^{2}+w_{p}^{2})}{w_{o,2}^{2} w_{p}^{2}+
w_{o,1}^{2}~(w_{o,2}^{2}+ w_{p}^{2})}
\end{equation}

\begin{figure}[tbp]
\par
\begin{center}
\includegraphics[angle=0, width=8cm, keepaspectratio=true]{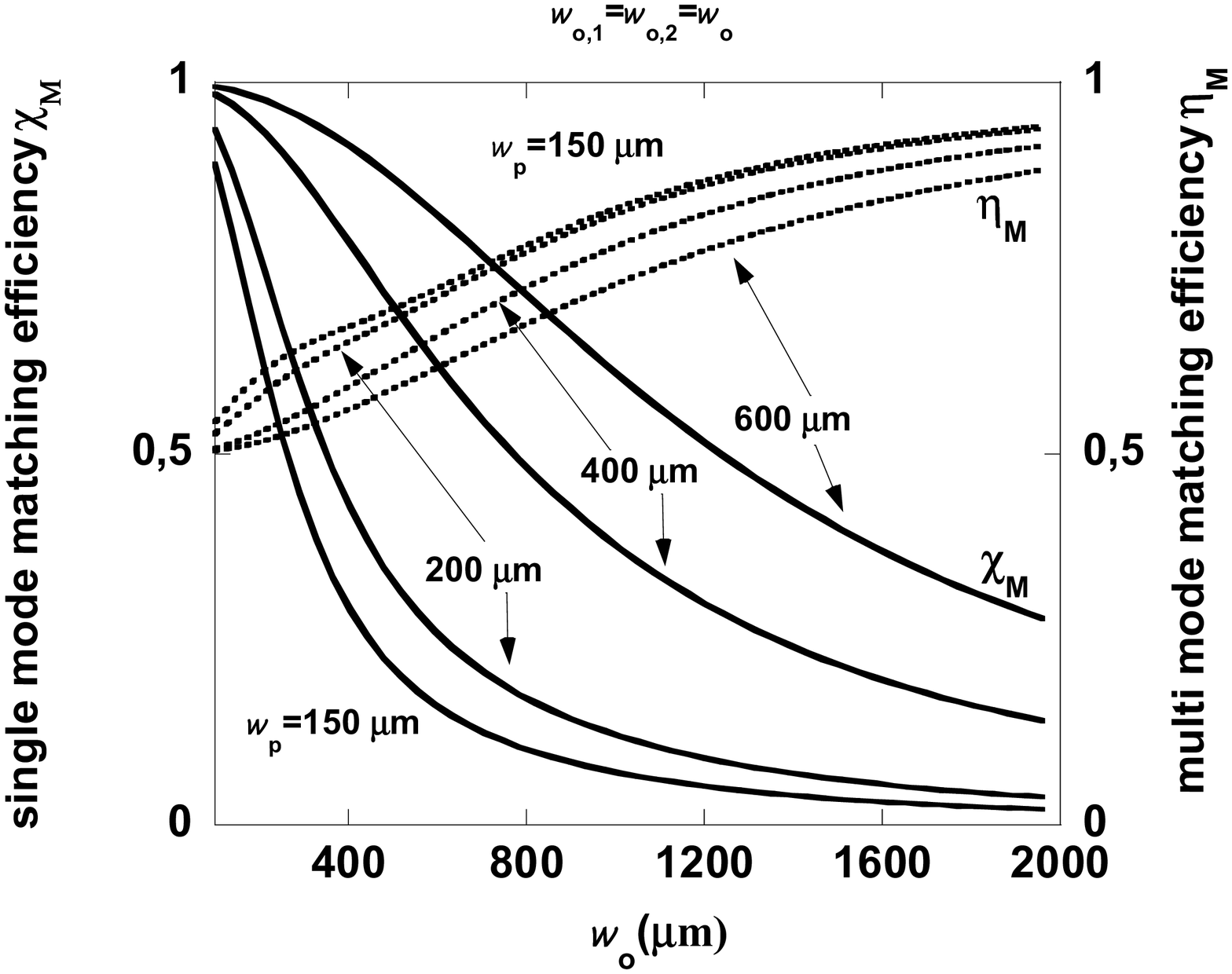}
\end{center}
\caption{Plots of $\eta_{M}$ (dashed) and $\chi_{M}$ (solid) for
$w_{o,1}=w_{o,2}=w_{o}$ (solid lines) and various pump waists,
$w_{p}=150, 200, 400, 600 ~\mu$m,  versus $w_{o}$.} \label{fig4}
\end{figure}

In the multi-mode approach presented here, the preparation and
collection modes can be thought of as spatially filtering or
selecting the multi-mode input light. As Fig.\ref{fig4}
demonstrates, the predictions made by this model yield different
results than the single-mode model.
The multi-mode model predicts that, for a fixed pump waist, the
maximum multi mode-matching efficiency is obtained when the
fiber-defined collection mode (at the crystal) is large, i.e. all
the pumped crystal volume is in a region of unit collection
efficiency of the spatial filter system. With the single-mode
model, one of the fibers acts as a source of a single-mode beam
that propagates through a spatial filter (in this case the pumped
crystal volume) to the other fiber. The maximum mode-matching
efficiency is achieved with a large pump waist, with respect to
the preparation and collection beam waist at the crystal. If the
pump waist is smaller than the fiber-defined collection beam
waist, the mode-matching efficiency is reduced. In this case,
 the efficiency is maximized with $w_{o,1}=w_{o,2}=w_{o}$,
but this presents the practical difficulty of having, and
aligning, exactly the same selected modes. Here, the differences
between the two models are more evident. In fact, when the
collection/preparation waist is much greater than the pump waist,
$\eta_{M}$ asymptotically goes to $1$, while $\chi_{M}$ goes to
$0$; in the opposite condition ($w_{p}\gg w_{o}$),
$\eta_{M}\rightarrow \frac{1}{2}$ and $\chi_{M}\rightarrow 1$.

\begin{figure}[tbp]
\par
\begin{center}
\includegraphics[angle=0, width=8cm,  keepaspectratio=true]{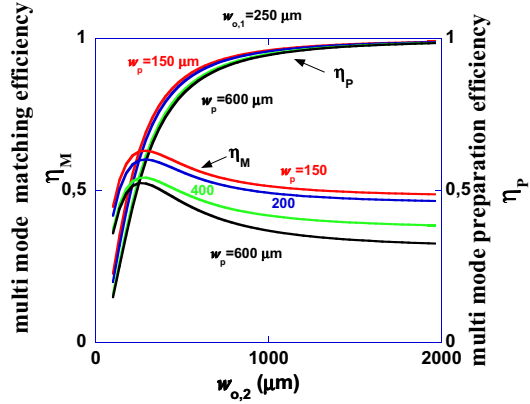}
\end{center}
\caption{Plot of $\eta_{p,M}$ versus the collecting mode waist for
various pump waist, $w_{p}=150, 200,400,600 \mu$m and for fixed
preparing waist $w_{o,1}=250$ $\mu$m. \label{fig5}}
\end{figure}

In Fig.\ref{fig5}, we plot the theoretical prediction of
$\eta_{P,M}$ for a 5 mm crystal of LiIO$_{3}$ versus the
collecting waist $w_{o,2}$ for fixed preparing waist $w_{o,1}=250$
 $\mu$m at the crystal with a range of pump waists. As it appears
for $\eta_{p}$ there is not an optimum collecting waist and the
highest efficiency is reached faster for the smaller pump waist.
The $\eta_M$  presents instead a finite optimum waist for smaller
pump waist, but the highest efficiency is never reached, because
modes are matched only on average, given the statistical nature of
the multimode model.

\section{Single mode preparation and multimode collection
efficiency} 
\begin{figure}[tbp]
\par
\begin{center}
\includegraphics[angle=0, width=12 cm, height=9 cm]{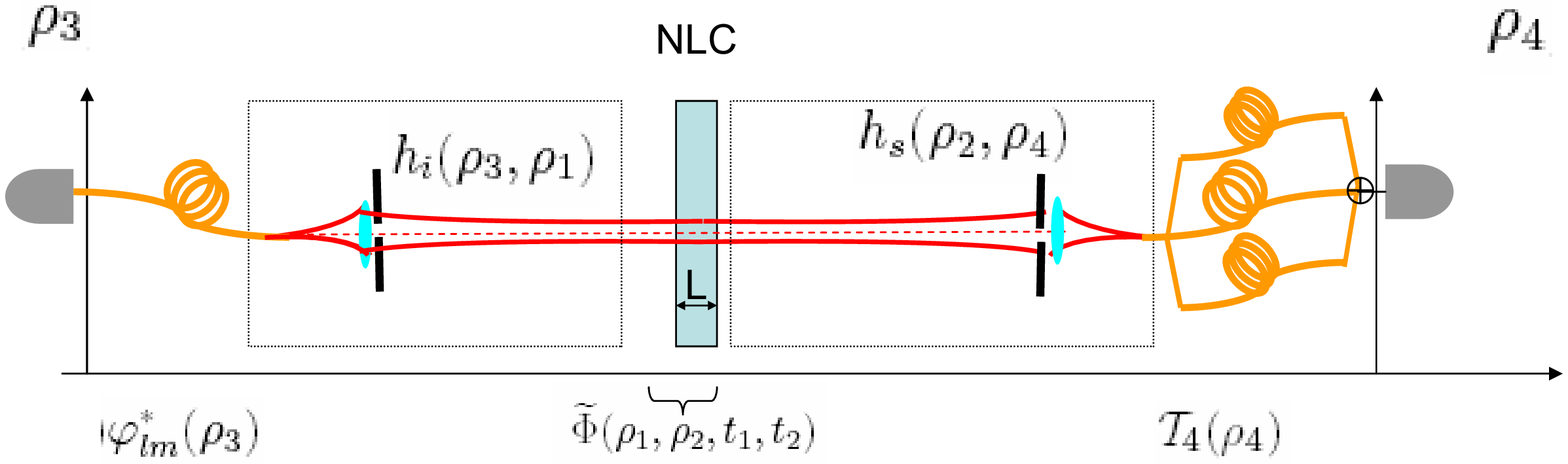}
\end{center}
\caption{Unfolded scheme  for the preparation of the heralding
channel in a single mode, while photons are collected in a bucket
detection scheme by an aperture or a multi-mode fiber placed in
the focus of a lens \label{fig6}}
\end{figure}

When we prepare the source according the unfolded
scheme of Fig.\ref{fig6} in a single mode, while we collect in a
multi-mode, the single counts are $\mathcal{C}_3$ in
Eq.(\ref{eq:3}), and the coincidences are:
\begin{equation}
      \mathrm{C}_{34}  = \int \textrm{d} t_1
       \textrm{d} t_2 \textrm{d}^2
    \rho_4
    \mathcal{T}(\rho_{4})  \left|   \int
    \textrm{d}^2
    \rho_1 \textrm{d}^2
    \rho_2 \textrm{d}^2
    \rho_3
    \widetilde{\Phi}_{12}( \mathbf{\rho}_1,t_{1}, \mathbf{\rho}_2,t_{2})
h_{i}(\mathbf{\rho}_1,\mathbf{\rho}_3)h_{s}(\mathbf{\rho}_2,\mathbf{\rho}_4)
\varphi^{*}_{lm} ( \rho_{3}) \right|^2
\end{equation}

\begin{figure}[tbp]
\par
\begin{center}
\includegraphics[angle=0, width=8cm,  keepaspectratio=true]{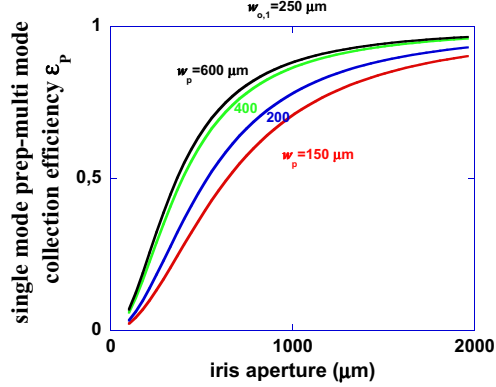}
\end{center}
\caption{Plot of $\epsilon_{P}$ versus the collecting iris
aperture for various pump waist, $w_{p}=150, 200,400,600 \mu$m
\label{fig7}}
\end{figure}

 The single-mode preparation and multi-mode collection efficiency is
\begin{equation}\label{chi}
   \epsilon_{P} = \frac{\mathrm{C}_{34}}{ \mathcal{C}_3  }.
\end{equation}
We explicitly  calculate it with a mode preparation given by a
single-mode fiber in a perfect imaging configuration (lens with
infinite aperture and magnification $M_3$ arranged to place the
collection beam waist, $w_{o,1}$ at the crystal), described as a
Gaussian field, $\varphi_{10} (\mathbf{\rho}_{3})$, while
collecting in a multi-mode with a Gaussian aperture $
\mathcal{T}_{4}(\mathbf{\rho}_{4})=e^{-\frac{2\rho_{4}^2 }{w^2}}.$
The signal optical system is $
h_s(\mathbf{\rho}_2,\mathbf{\rho}_4)=\exp[-i  k
(\mathbf{\rho}_2-\mathbf{\rho}_4)^2)] $, the free space
propagation, with $k=\pi /(\lambda_{s} d)$ and $d$ the propagation
distance. The single mode preparation and multi-mode collection
efficiency is given by
\begin{equation}\label{epsilon}
\epsilon_{P}=\mathcal{F'}_P \frac{~ (-\mathcal{N}_{p}+
\mathcal{N}_{s}+ K_{p}\theta_{i})k^2 w^2 w_{o,1}^2
w_{p}^{2}~\sqrt{w_{o,1}^2+w_{p}^2}}{\sqrt{C' B'}},
\end{equation}
with
\begin{equation}\label{F} \mathcal{F'}_P=
 \frac{ \mathrm{Erf}[ \frac{L \sqrt{2 B'}}{K_{p} \sqrt{C'}}]}{\mathrm{Erf}[\sqrt{2}
L\frac{(-\mathcal{N}_{p}+ \mathcal{N}_{s}+ K_{p}
\theta_{i})}{K_{p} \sqrt{w_{o,1}^{2}+ w_{p}^{2}}}]},
\end{equation}
 $B'=( \mathcal{N}_{p}-  \mathcal{N}_{s} +K_{p}\theta_{s})^2
w_{o,1}^4 w_p^2 k^2+ (- \mathcal{N}_{p}+ \mathcal{N}_{s}+
K_{p}\theta_{i})^2 (w_{o,1}^2+w_p^2+k^2 w^2 w_{o,1}^2 w_p^2)+
K_{p}^{2} k^2 w_{o,1}^2 w_p^4 (\theta_{i}+\theta_{s})^2 $, and
$C'=(w_{o,1}^2+w_p^2)^2+k^2 w_{o,1}^2 w_p^2 (w^2 w_p^2 +w_{o,1}^2
w^2+ w_{o,1}^2 w_p^2 )$.
 In the
thin crystal limit this reduces to
\begin{equation}\label{epsilon0}
\epsilon_{P}= \frac{~k^2 w^2 w_{o,1}^2 w_{p}^{2} (w_{o,1}^2 +
w_{p}^2 )}{(w_{o,1}^2+w_p^2)^2+k^2 w_{o,1}^2 w_p^2 (w^2 w_p^2
+w_{o,1}^2 w^2+ w_{o,1}^2 w_p^2 ) },
\end{equation}
We observed that in this case $\epsilon_P$ depends slightly on
$L$. We plot in Fig.\ref{fig7},the theoretical prediction of
$\epsilon_P$ for a 5 mm crystal of LiIO$_{3}$ versus the iris
aperture $w$ for fixed preparing waist $w_{o,1}=250 \mu$m at the
crystal, for different pump waist configurations. In this case we
have an intermediate behavior between $\chi_{P}$ and $\eta_ {P}$,
the highest efficiency is obtained faster for larger pump waist.
We underline that the comparison with the following experimental
results show a qualitative analogy with this curve, because the
$k$ parameter plays an essential rule in the estimate of the
efficiency. Specifically the quantitative value of $k$ in the
simulation does not correspond to the experimental values.

\section{Preliminary experimental results} We give here some preliminary
experimental results and compare them with theoretical
predictions, showing the agreement. The simulations and
measurements are done exciting PDC in a type I phase-matching
configuration with a pump wavelength of 351 nm and by a LiIO$_{3}$
crystal length of 5 mm. Pump waists at the crystal were ranging
from from $150 ~\mu$m to $600 ~\mu$m.

Some measurements were done with the trigger arm (heralding
channel) coupled with a single-mode fiber. The lens images the
minimum waist at the crystal. The signal arm was either coupled
with a single-mode fiber in a perfect imaging configuration or
with a multimode-fiber placed in the focus of the coupling lens,
where an iris at the lens selects the collecting modes. Other
measurements were performed in a similar setup (6 mm LiIO$_{3}$)
at the National Institute of Standards and Technology (NIST) with
the trigger beam coupled with a single-mode fiber and the signal
coupled to the detector by a single lens, with an iris selecting
the modes. In particular two experimental configurations were
chosen. In one configuration the trigger mode was fixed to
maximize the singles rate, while in the other configuration the
heralding channel waist was matched to the pump waist at the
crystal.

\begin{figure}[tbp]
\par
\begin{center}
\includegraphics[angle=0,width=8cm,keepaspectratio=true]{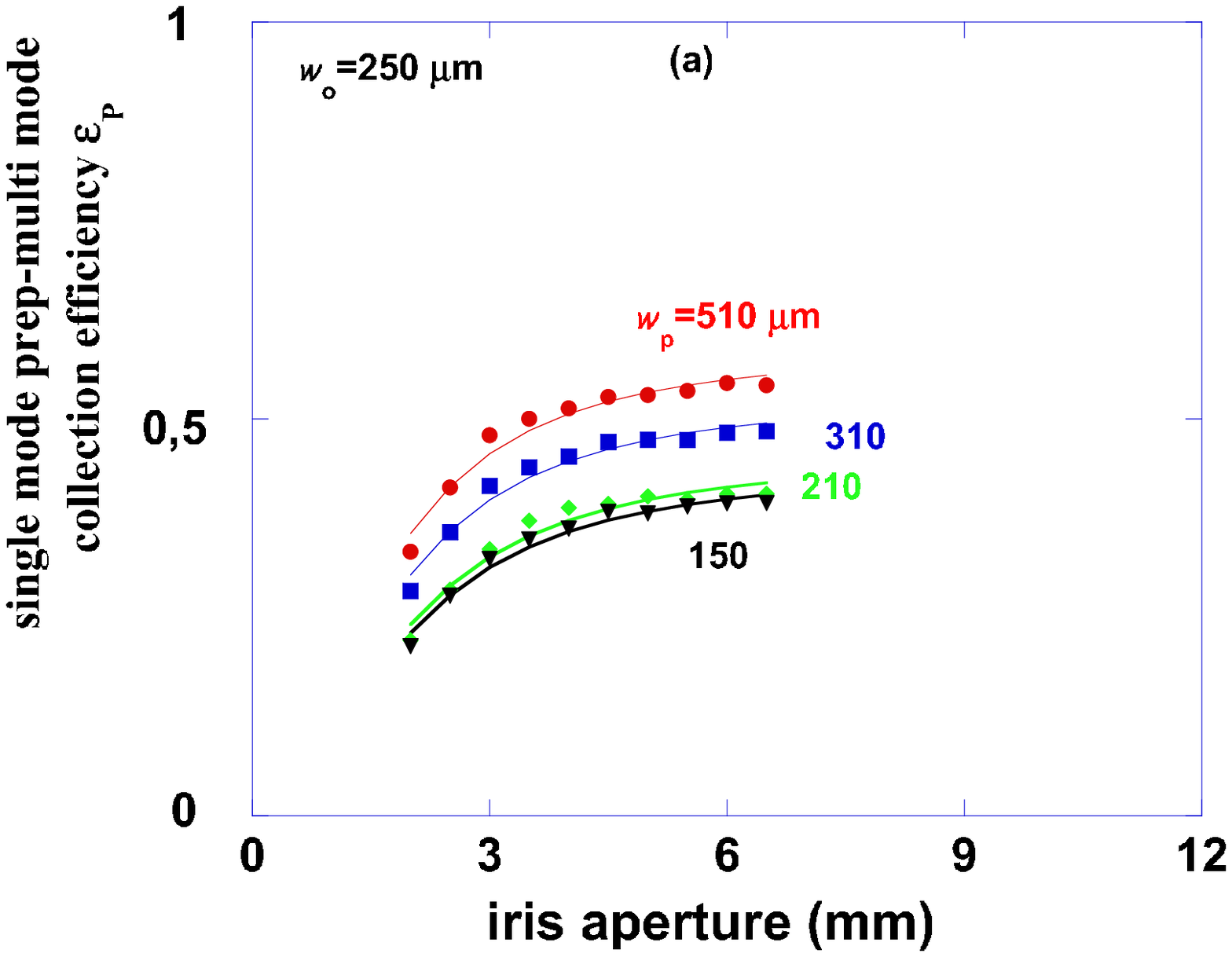}
\includegraphics[angle=0,width=8cm,keepaspectratio=true]{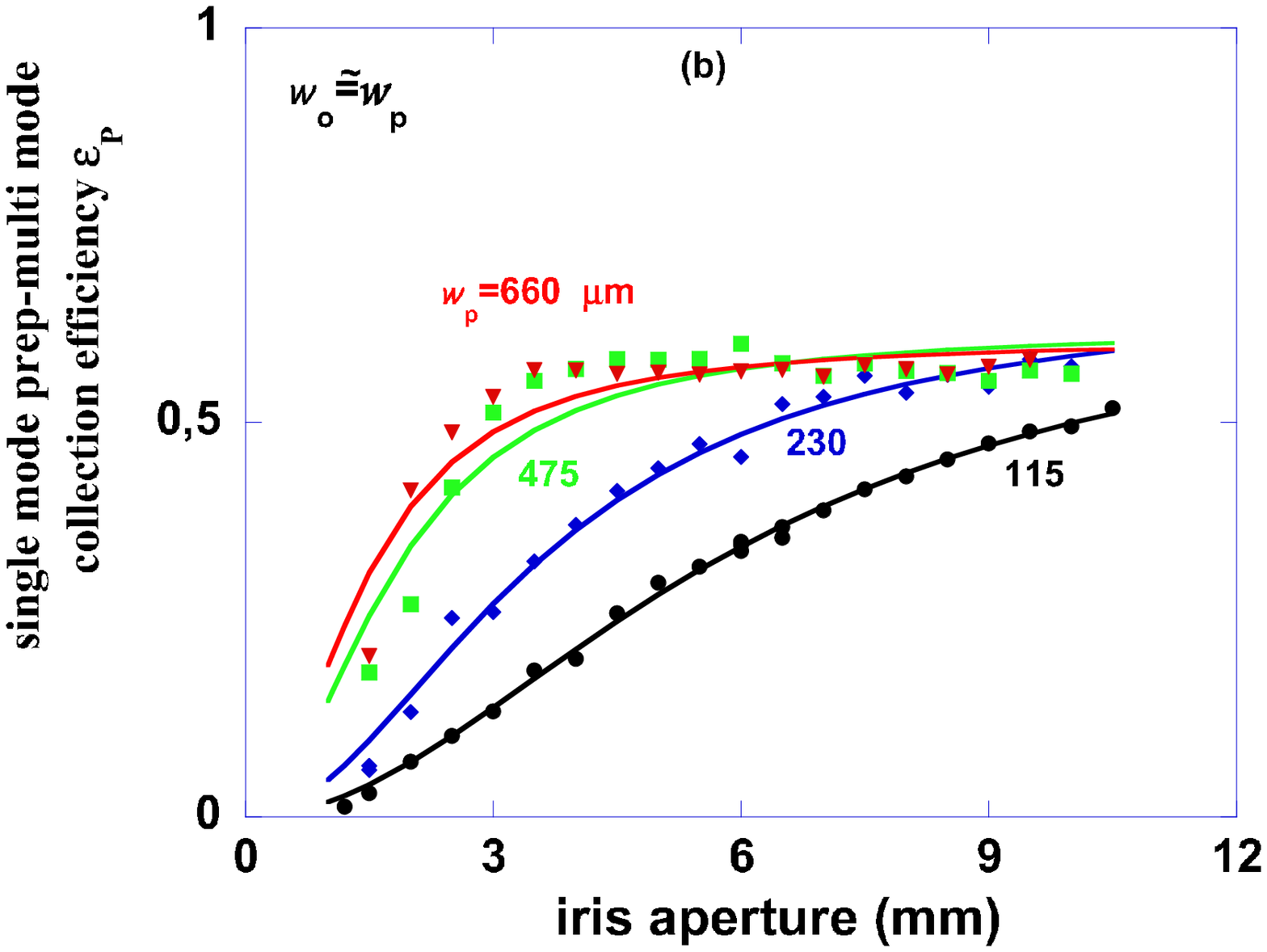}
\end{center}
\caption{Plots of $\epsilon_{P}$ data versus the iris diameter (a)
for a range of pump waist with fixed $w_{o}=250$ $\mu$m   and (b)
for a range of pump waists with $w_{p}\approx w_{o}$. (a) shows
measurements performed at IEN, where the efficiency is uncorrected
for the detector dead-time, while (b) shows measurements performed
at NIST, corrected for dead-time. The solid line are fits obtained
with the expected theoretical curve in the thin-crystal limit
reported in Eq.(\ref{epsilon0}).} \label{fig8ab}
\end{figure}

In Fig. \ref{fig8ab} we show the experimental values of the
single-mode preparation and multi-mode collection efficiency
($\epsilon_P$) versus the collecting iris diameter, for the two
experimental configurations, i.e. for fixed $w_{o}=250$ $\mu$m
 and (b) for fixed  $w_{p}\approx w_{o}$ and a
range of pump waists as indicated. In both cases the maximum
efficiency is obtained for larger pump waists as predicted by the
theory in Fig. \ref{fig7}.

We point out that the theoretical prediction of the multi-mode
collection and multi mode preparation efficiency ($\eta_P$) versus
the equivalent Gaussian filter collection aperture $w_{o,2}$ would
instead give the maximum efficiency for smaller pump waists.
However, while the single-mode preparation and matching efficiency
($\chi_{M,P}$) versus the collecting waist $w_{o,2}$ present a
maximum for properly matching the three modes, $\eta_{P,M}$ may
not reach the highest efficiency even in this case.

\begin{figure}[tbp]
\par
\begin{center}
\includegraphics[angle=0, width=10cm,keepaspectratio=true]{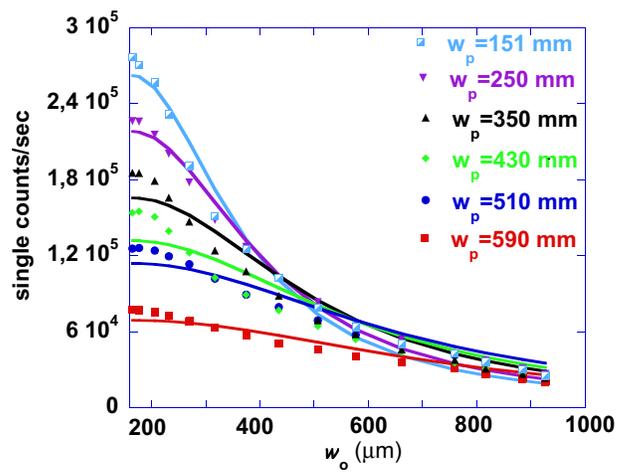}
\end{center}
\caption{Plots of single counts $\mathcal{C}_{3}$ versus
 $w_{o}$  for a range of pump waists. The maximum counts are for the smallest pump waist.}
  \label{fig9}
\end{figure}

\begin{figure}[tbp]
\par
\begin{center}
\includegraphics[angle=0, width=10cm,keepaspectratio=true]{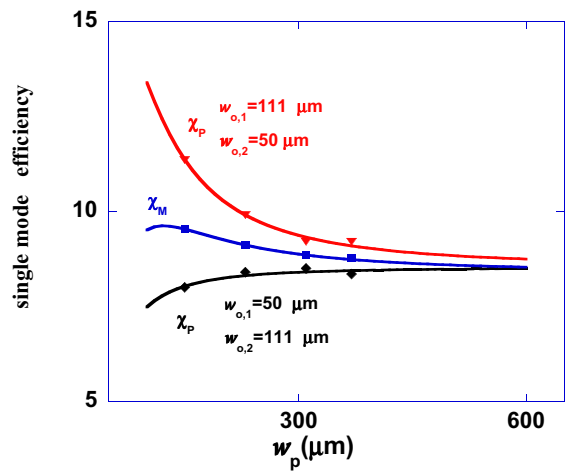}
\end{center}
\caption{Plots of the single mode preparation efficiency
$\chi_{P}$  and matching efficiency  $\chi_{M}$ versus the pump
waist $w_{p}$ for two preparing and collecting mode
configurations,
 $w_{o,1}=111~\mu m$ and $w_{o,2}=50~\mu
m$ the upper curve and $w_{o,1}=50~\mu m$ and $w_{o,2}=111~\mu m$
the lower curve. The curve in the middle is $\chi_{M}$ for
$w_{o,1}=155~\mu m$ and $w_{o,2}=70~\mu$. The solid lines are fit
obtained by the proposed theoretical model.}
  \label{fig10}
\end{figure}

Fig. \ref{fig9} shows the experimental single counts from a
heralding channel coupled with a single-mode fiber versus the
fiber waist at the crystal, for various pump waists. Data are
fitted by theoretical curves that for single mode coupling are
$\mathcal{C}_3=\frac{K_{p} Erf[\frac{L \sqrt{2}
(-N_{p}+N_{s}+K_{p} \theta_{i}) }{L K_p (\sqrt{w_{o}^{2}+
w_{p}^2})}]} {\sqrt{2 \pi}(-N_{p}+N_{s}+K_{p}
\theta_{i})(\sqrt{w_{o}^{2}+ w_{p}^2} )}$; here the data clearly
matches the single-mode propagation because we have an optimum
preparation waist that maximizes singles rate (note that the
maximum singles rate are for a smaller pump waist). In a
multi-mode configuration the single rates would just increase with
the iris or filter aperture.
In Fig. \ref{fig10} we plot the experimental data of the
uncorrected efficiency in the case of single mode fiber placed on
both detection arms for a 5 mm LiIO$_3$ versus pump waist at the
crystal. Measurements were performed for two different preparing
and collecting mode waists configuration, that is the apparent
waists at the crystal were $w_{o,1}=110~\mu m$ and $w_{o,2}=50~\mu
m$ and reversed. The data are fitted by the Eq.(\ref{chiP0}) and
Eq.(\ref{chiM0}), in the thin crystal configuration, this is
because of the very small size of the waists, where the theory for
long crystal is no longer reliable. The agreement shows the
correctness of model proposed.

\section{Conclusions}
In summary, we have presented an analytic model to quantify the
mode preparation and matching efficiency in terms of adjustable
experimental parameters with the goal of optimizing single-mode
collection from PDC sources. In addition, we have presented an
alternative scheme that may have more validity for multi-mode
collection arrangements. We reported preliminary experimental
results, supporting the validity of the proposed multi-mode
collection model.

\section{Acknowledgments}
This work was supported in part by DARPA/QUIST, ARDA, ARO.

\bibliography{singlepho1}   
\bibliographystyle{spiebib}   

\end{document}